\documentclass[a4paper]{article}
\usepackage[ansinew]{inputenc}
\usepackage{times}
\usepackage[T1]{fontenc}
\usepackage{graphicx}
\usepackage{geometry}
\usepackage{amssymb}
\usepackage{amsmath}

\usepackage{rotating}

\usepackage{xcolor}
\colorlet{shadecolor}{gray!25}
\usepackage{subfig}
\usepackage{float}
\usepackage{lscape}

\author{Ludwig A. Hothorn,\\ 
Im Grund 12, D-31867 Lauenau, Germany\\ \scriptsize(retired from Leibniz University Hannover)}

\title{Statistical evaluation of \textit{in-vivo} bioassays in regulatory toxicology considering males and females}
\begin{document}

\maketitle
\begin{abstract}

The separate evaluation for males and females is the recent standard in in-vivo toxicology for dose or treatment effects using Dunnett
tests. The alternative pre-test for sex-by-treatment interaction is problematic. Here a joint test is proposed considering the two sex-specific and the pooled Dunnett-type comparisons. The calculation of either simultaneous confidence intervals or adjusted p-values with the R-package multcomp is demonstrated using a real data example.
\end{abstract}

\section{The problem}\label{sec1}
Almost all \textit{in-vivo} bioassays in regulatory toxicology use a design where multiple dose (treatment) groups are compared with a concurrent negative control, e.g. selected clinical chemistry endpoints were compared  three dose groups with disodium adenosine-triphosphate with a water control \cite{Jager2020}. Typically, the Dunnett procedure is used \cite{Dunnett1955} and proposed in related guidance's \cite{NTP2012}.
Almost all such \textit{in-vivo} bioassays use both males and females, and evaluate almost all for males and females separately, each at level $\alpha$. The advantage is simplicity. The disadvantage are neither statistical claim for no sex-specificity, nor for particular sex-specific effects and especially the loss of power because considering $N/2$ each if there is no significant sex effect as well as then findings to be interpreted twice.\\
In addition to the separate analysis described above, a stepwise approach can be found in the literature for such treatment-by-sex two-way designs: if the pre-F-test for treatment-by-sex interaction is significant, analysis is performed separately; otherwise pooled over both sexes. This approach is not really suitable for the evaluation of in-vivo assays \cite{Hothorn2020l}.

\section{A motivating data example}

A motivating data example, the relative liver weight data from a 13-week study on F344 rats administered with sodium dichromate dihydrate is used here \cite{NTP}. The boxplot reveals an interaction between males and females and possible weight reduction in males as well as approximate homogeneous variance in symmetric distributed data. I.e. the standard Dunnett-test using the R-package multcomp \cite{Hothorn2008} is recommended.

\begin{figure}[htbp]
	\centering
		\includegraphics[width=0.450\textwidth]{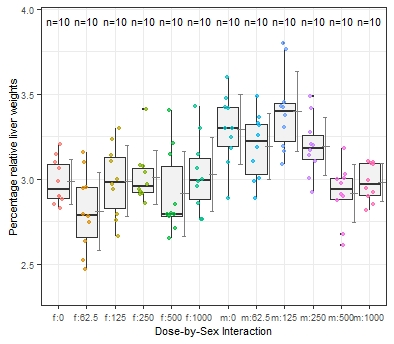}
	\caption{Relative liver weights}
	\label{fig:LiverIA}
\end{figure}

The ANOVA analysis reveal clear dose, sex and interaction effects, i.e. a separate analysis for males and females is indicated according to the recent paradigm.

\begin{table}[ht]
\centering\footnotesize
\begin{tabular}{lrrrrr}
  \hline
 & Df & Sum Sq & Mean Sq & F value & Pr($>$F) \\ 
  \hline
	Dose & 5 & 1.07 & 0.21 & 5.68 & 0.0001 \\ 
  Sex & 1 & 1.27 & 1.27 & 33.61 & 0.0000 \\ 
  Dose:Sex & 5 & 0.95 & 0.19 & 4.99 & 0.0004 \\ 
  Residuals & 108 & 4.09 & 0.04 &  &  \\ 
   \hline
\end{tabular}
\caption{Global ANOVA} 
\end{table}

\section{Simultaneous Dunnett-type comparisons in two-way designs}
A pre-test for no interaction is problematic, because first of all it should be an equivalence test, where the definition of the still to be tolerated threshold for an interaction is difficult to practically impossible. And second, neither the error control nor the power of the conditional joint procedure (pre-test, main test) is trivial and satisfactory. Since the inference for the factor dose is primary and the secondary factor has only a few levels (here for sex exactly 2), a simultaneous single-step analysis, considering two Dunnett tests jointly, one for males and the other for females, can be recommended instead of a pre-test procedure \cite{Hothorn2020t}. Here we recommend the further jointly inference of the data pooled over both sexes. This allows the interpretation of both global and sex-specific effects. Due to the high correlation between these three test statistics, the additional price of conservatism, i.e. increase in the false negative rate, is still acceptable compared to the advantages of a consistent interpretation \cite{Hothorn2020l}. This approach requires the computation of the more complex correlation matrix in the package multcomp for a pseudo one-way layout, the so-called cell means model without intercept \cite{Kitsche2015}. Both simultaneous confidence intervals (preferred because of \cite{Hothorn2020} ) and adjusted p-values are then easily available. In the next chapter this approach is illustrated by the data example.

\section{Evaluation of the data example using library(multcomp)}
The formulation of  the cell means model is simple as a one-way layout for the interaction factor $\text{Dose:Sex}$. A related simple linear model is fitted for this interaction factor without intercept. For the function \textit{glht()} in the R-library multcomp, the contrast matrix for the joint test is formulated to (see the R-code in the Appendix):

\scriptsize
\begin{verbatim}
              0 62.5 125 250 500 1000  0 62.5 125 250 500 1000                        
f:62.5 - 0 -1.0  1.0 0.0 0.0 0.0  0.0  0.0 0.0 0.0 0.0 0.0 0.0
f:125 - 0  -1.0  0.0 1.0 0.0 0.0  0.0  0.0 0.0 0.0 0.0 0.0 0.0
f:250 - 0  -1.0  0.0 0.0 1.0 0.0  0.0  0.0 0.0 0.0 0.0 0.0 0.0
f:500 - 0  -1.0  0.0 0.0 0.0 1.0  0.0  0.0 0.0 0.0 0.0 0.0 0.0
f:1000 - 0 -1.0  0.0 0.0 0.0 0.0  1.0  0.0 0.0 0.0 0.0 0.0 0.0
m:62.5 - 0  0.0  0.0 0.0 0.0 0.0  0.0 -1.0 1.0 0.0 0.0 0.0 0.0
m:125 - 0   0.0  0.0 0.0 0.0 0.0  0.0 -1.0 0.0 1.0 0.0 0.0 0.0
m:250 - 0   0.0  0.0 0.0 0.0 0.0  0.0 -1.0 0.0 0.0 1.0 0.0 0.0
m:500 - 0   0.0  0.0 0.0 0.0 0.0  0.0 -1.0 0.0 0.0 0.0 1.0 0.0
m:1000 - 0  0.0  0.0 0.0 0.0 0.0  0.0 -1.0 0.0 0.0 0.0 0.0 1.0
p:62-0     -0.5  0.5 0.0 0.0 0.0  0.0 -0.5 0.5 0.0 0.0 0.0 0.0
p:125-0    -0.5  0.0 0.5 0.0 0.0  0.0 -0.5 0.0 0.5 0.0 0.0 0.0
p:250-0    -0.5  0.0 0.0 0.5 0.0  0.0 -0.5 0.0 0.0 0.5 0.0 0.0
p:500-0    -0.5  0.0 0.0 0.0 0.5  0.0 -0.5 0.0 0.0 0.0 0.5 0.0
p:1000-0   -0.5  0.0 0.0 0.0 0.0  0.5 -0.5 0.0 0.0 0.0 0.0 0.5
\end{verbatim}
\normalsize
Instead of a $(k * (k+1)$ matrix for the one-way layout a $(3k * 2(k+1)$ matrix is used for the cell means model, where the upper parts are for the comparisons with males (m) or females (f) and the lower part for the pooled (p). 

\begin{table}[ht]
\centering\footnotesize
\begin{tabular}{rlr}
  \hline
Type & Comparison & p-value \\ 
  \hline
  Females & 62.5-0 & 0.34 \\ 
  & 125-0 & 0.99 \\ 
  & 250-0 & 0.99 \\ 
  & 500-0 & 0.99 \\ 
  & 1000-0 & 0.99 \\ \hline
  Males & 62.5-0 & 0.91 \\ 
  & 125-0 & 0.87 \\ 
  & 250-0 & 0.91 \\ 
   & 500-0 & 0.0004 $\Downarrow$ \\ 
   & 1000-0 & 0.007 $\Downarrow$ \\ \hline
  Pooled &62.5-0 & 0.23 \\ 
   & 125-0 & 0.98 \\ 
  & 250-0 & 0.99 \\ 
   & 500-0 & 0.005 $\Downarrow$\\ 
   & 1000-0 & 0.25 \\ 
  \hline
\end{tabular}
\caption{Adjusted two-sided p-values for the joint test} 
\end{table}
\normalsize
In this example the global residual $df$ is with  108 large, therefore the $df$-reduction in separate analysis plays a smaller role than the usual designs with smaller $n_i$. Therefore, the p-value for the $[1000-0]$ comparison of $p^{\text{males only}}=0.0019$ is relatively much smaller compared to the joint test of $p^{\text{joint test}}=0.0064$.\\
The simultaneous two-sided confidence intervals in Figure 2 allow a qualitative interpretation in the percentage scale of the relative liver weights.
\begin{figure}[htbp]
	\centering
		\includegraphics[width=0.50\textwidth]{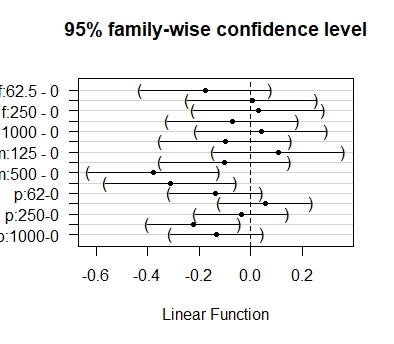}
	\caption{Joint test: simultaneous confidence intervals for the difference to control}
	\label{fig:CIliv}
\end{figure}

For a toxicological view, a liver weight reduction in the two high doses (1000,500) can be selectively detected in males only. The pooled analysis is biased.\\

\section{Conclusions}
For the evaluation of possible sex-specific effects in \textit{in-vivo} toxicology of joint Dunnett-type test considering males, females and pooled analysis by means of a combined contrast matrix. Simultaneous confidence intervals or adjusted p-values are available by means of the R package multcomp. Interpretation of possible compound-related effects is straightforward.\\
Generalizations for counts, proportions and poly-k estimates in the generalized linear model as well as for non-parametric models \cite{Konietschke2012} and variance heterogeneity will be presented soon. Furthermore, this joint test will soon be available for the Williams-Trend test to examine monotonous effects.

\bibliographystyle{plain}
\footnotesize


\section{Appendix: R-Code of the example}
\scriptsize
\begin{verbatim}
livMF <-
  structure(list(Dose = c(0, 0, 0, 0, 0, 0, 0, 0, 0, 0, 62.5, 62.5,
                          62.5, 62.5, 62.5, 62.5, 62.5, 62.5, 62.5, 62.5, 125, 125, 125,
                          125, 125, 125, 125, 125, 125, 125, 250, 250, 250, 250, 250, 250,
                          250, 250, 250, 250, 500, 500, 500, 500, 500, 500, 500, 500, 500,
                          500, 1000, 1000, 1000, 1000, 1000, 1000, 1000, 1000, 1000, 1000,
                          0, 0, 0, 0, 0, 0, 0, 0, 0, 0, 62.5, 62.5, 62.5, 62.5, 62.5, 62.5,
                          62.5, 62.5, 62.5, 62.5, 125, 125, 125, 125, 125, 125, 125, 125,
                          125, 125, 250, 250, 250, 250, 250, 250, 250, 250, 250, 250, 500,
                          500, 500, 500, 500, 500, 500, 500, 500, 500, 1000, 1000, 1000,
                          1000, 1000, 1000, 1000, 1000, 1000, 1000),
                 BodyWt = c(337.6,319.2, 368.7, 373.3, 314.9, 306, 326.2, 331.7, 319.7, 304.9,
                            321, 325.9, 342.9, 322.6, 330.6, 302.2, 325.2, 326.7, 297.1,
                            329.8, 350.8, 305.5, 333.3, 348.2, 359.9, 345.7, 338, 322.2,
                            351.2, 309.8, 338, 337.3, 338.8, 329.7, 349.2, 298.4, 332.6,
                            323.3, 318.3, 330.6, 306.1, 320.3, 301.1, 320.7, 326.8, 313.4,
                            324.8, 315.2, 320.1, 308.2, 278, 297.8, 320.9, 313.4, 291.2,
                            293.8, 294.4, 280.9, 316.8, 292.2, 211.2, 199.4, 199.3, 179.8,
                            183.3, 180.2, 192.4, 186.4, 201.9, 193.5, 209.9, 212.2, 212.1,
                            211.8, 217.3, 203.1, 208.7, 243, 215.9, 215.1, 188.9, 202.6,
                            209.1, 190.5, 192, 210, 199.6, 194.2, 205.1, 201.2, 195.8, 205.7,
                            191.6, 199, 199, 184.8, 204.8, 207.4, 203.6, 193.1, 182.5, 191.6,
                            192.7, 196.2, 176.7, 196.4, 209.7, 199.1, 187.1, 196.5, 192.7,
                            187, 177.7, 184.2, 182.2, 178.3, 193, 180.3, 186.4, 186),
                 LiverWt = c(10.73,9.89, 12.6, 13.43, 10.21, 10.64, 10.76, 11.34, 10.55, 8.79, 9.27,
                             10.62, 11.53, 9.68, 10.53, 10.05, 10.09, 10.84, 8.87, 11.49,
                             12.02, 9.42, 12.55, 11.11, 13.67, 10.94, 11.66, 11.03, 11.85,
                             10.24, 10.7, 10.59, 9.89, 10.57, 10.84, 9.54, 11.35, 10.58, 11.09,
                             9.93, 9.2, 9.52, 9.58, 9.62, 9.89, 8.18, 9.41, 9.05, 8.56, 8.96,
                             7.92, 9.25, 9.38, 9.23, 9.02, 9.07, 9.1, 7.9, 9.48, 8.46, 6.76,
                             5.78, 5.63, 5.51, 5.76, 5.22, 5.54, 5.56, 5.76, 5.99, 6.5, 6.25,
                             5.89, 6.68, 6.41, 5.67, 5.26, 6.39, 5.33, 5.9, 5.93, 6.01, 6.89,
                             5.71, 6.21, 6.17, 6.19, 5.35, 5.72, 5.36, 6.68, 6, 5.61, 5.91,
                             6.13, 5.43, 6.29, 6.03, 6.19, 5.51, 5.86, 5.35, 5.22, 5.48, 4.94,
                             5.2, 6.6, 6.78, 5.2, 5.6, 6.35, 5.6, 4.91, 5.79, 5.44, 5.09,
                             6.62, 4.98, 5.51, 5.7),
                 Sex = structure(c(2L, 2L, 2L, 2L, 2L,2L, 2L, 2L, 2L, 2L, 2L, 2L, 2L, 2L, 2L, 2L, 2L, 2L, 2L, 2L, 2L,
                          2L, 2L, 2L, 2L, 2L, 2L, 2L, 2L, 2L, 2L, 2L, 2L, 2L, 2L, 2L, 2L,
                          2L, 2L, 2L, 2L, 2L, 2L, 2L, 2L, 2L, 2L, 2L, 2L, 2L, 2L, 2L, 2L,
                          2L, 2L, 2L, 2L, 2L, 2L, 2L, 1L, 1L, 1L, 1L, 1L, 1L, 1L, 1L, 1L,
                          1L, 1L, 1L, 1L, 1L, 1L, 1L, 1L, 1L, 1L, 1L, 1L, 1L, 1L, 1L, 1L,
                          1L, 1L, 1L, 1L, 1L, 1L, 1L, 1L, 1L, 1L, 1L, 1L, 1L, 1L, 1L, 1L,
                          1L, 1L, 1L, 1L, 1L, 1L, 1L, 1L, 1L, 1L, 1L, 1L, 1L, 1L, 1L, 1L,
                          1L, 1L, 1L),
                .Label = c("f", "m"), class = "factor")), row.names = c(NA,120L), class = "data.frame")

livmf <-livMF
livmf$relLiv <-livmf$LiverWt/livmf$BodyWt
livmf$RelLiv <-100*livmf$relLiv 
livmf$Dose <-as.factor(livmf$Dose)

#######################################
library(multcomp)
livmf$ww <- with(livmf, interaction(Dose,Sex))
Du <- contrMat(table(livmf$Dose), "Dunnett")
K1 <- cbind(Du, matrix(0, nrow = nrow(Du), ncol = ncol(Du)))
rownames(K1) <- paste(levels(livmf$Sex)[1], rownames(K1), sep = ":")
K2 <- cbind(matrix(0, nrow = nrow(Du), ncol = ncol(Du)), Du)
rownames(K2) <- paste(levels(livmf$Sex)[2], rownames(K2), sep = ":")
K <- rbind(K1, K2)
colnames(K) <- c(colnames(Du), colnames(Du))
livmf$ww <- with(livmf, interaction(Dose,Sex))
Mod6 <- lm(RelLiv ~ ww - 1, data = livmf)
X6 <- glht(Mod6, linfct = K)
K3<-rbind("p:62.5-0"   = c(-1/2, 1/2,0,  0,  0,  0, -1/2, 1/2, 0,  0,  0,0),
          "p:125-0"    = c(-1/2, 0,  1/2,0,  0,  0, -1/2, 0,   1/2,0,  0,0),
          "p:250-0"    = c(-1/2, 0,  0, 1/2, 0,  0, -1/2, 0,   0,  1/2,0, 0),
          "p:500-0"    = c(-1/2, 0,  0,  0,  1/2,0, -1/2, 0,   0,  0,  1/2,0),
          "p:1000-0"   = c(-1/2, 0,  0,  0,  0, 1/2,-1/2, 0,   0,  0,  0, 1/2))
X7<-glht(Mod6, linfct = K7)
xx7<-fortify(summary(X7))
print(xtable(xx7[,c(1,6)], caption="Joint test"), digits=6)
\end{verbatim}


\begin{thebibliography}{1}

\bibitem{NTP}
13 weeks gavage study on f344 rats administered with sodium dichromate dihydrate
  (vi) (casrn: 7789-12-0, study number: C20114). technical report.
\newblock Technical report, US-NTP, 2007.

\bibitem{NTP2012}
National toxicology program. testing information, statistical procedures,
  expanded overview\\
  (http://ntp.niehs.nih.gov/?objectid=72015e2c-bdb7-ceba-f17f9aca7ae5346d).
\newblock Technical report, US-NTP, 2012.

\bibitem{Dunnett1955}
C.~W. Dunnett.
\newblock A multiple comparison procedure for comparing several treatments with
  a control.
\newblock {\em Journal of the American Statistical Association},
  50(272):1096--1121, 1955.

\bibitem{Hothorn2020}
L.~A. Hothorn and R.~Pirow.
\newblock Use compatibility intervals in regulatory toxicology.
\newblock {\em Regulatory Toxicology and Pharmacology}, 116:104720, October
  2020.

\bibitem{Hothorn2020l}
L.A. Hothorn.
\newblock Dunnett test in factorial designs. 
\newblock {\em In preparation} 2020.

\bibitem{Hothorn2020t}
T.~Hothorn.
\newblock Additional multcomp examples.
\newblock {\em Vignette to multcomp package} 2020.

\bibitem{Hothorn2008}
T.~Hothorn, F.~Bretz, and P.~Westfall.
\newblock Simultaneous inference in general parametric models.
\newblock {\em  Biometrical J}, 2008.

\bibitem{Jager2020}
R.~Jager, M.~Purpura, and J.~C. Fuller.
\newblock Subchronic (90-day) repeated dose toxicity study of disodium
  adenosine-5 '-triphosphate in rats.
\newblock {\em Regulatory Toxicology and Pharmacology}, 116:104760, October
  2020.

\bibitem{Kitsche2015}
A.~Kitsche and F.~Schaarschmidt.
\newblock Analysis of statistical interactions in factorial experiments.
\newblock {\em Journal of Agronomy and Crop Science}, 201(1):69--79, February
  2015.
	
	\bibitem{Konietschke2012}
F.~Konietschke and L.~A. Hothorn.
\newblock Rank-based multiple test procedures and simultaneous confidence
  intervals.
\newblock {\em Electronic Journal of Statistics}, 6:738--759, 2012.

\end{thebibliography}


\normalsize

\end{document}